\documentclass[lettersize,journal]{IEEEtran}
\usepackage{amsmath,amsfonts}
\usepackage{algorithmic}
\usepackage{algorithm}
\usepackage{array}
\usepackage[caption=false,font=normalsize,labelfont=sf,textfont=sf]{subfig}
\usepackage{textcomp}
\usepackage{stfloats}
\usepackage{url}
\usepackage{verbatim}
\usepackage{graphicx}
\usepackage{cite}
\usepackage{amssymb}
\usepackage{amsthm}
\usepackage{titlesec}
\usepackage{hyperref}
\usepackage{subcaption}
\usepackage{placeins}
\usepackage{float}
\usepackage{xcolor}
\usepackage{setspace}
\allowdisplaybreaks[4]
\usepackage[font=footnotesize, skip=3pt]{caption}
\usepackage[a4paper, left= 0.585in, right= 0.585in, top= 0.55in, bottom= 0.55in]{geometry}
\theoremstyle{remark}
\newtheorem*{theorem*}{Theorem}
\begin{document}
\vspace{-1.0cm}
\title{\LARGE{Rotatable Antenna-Enabled Mobile Edge Computing}}

\vspace{-0.6cm}
\author{{\fontsize{10pt}{\baselineskip}\selectfont
    {Qiyao Wang, Beixiong Zheng, {\em Senior Member, IEEE}, Xue Xiong, Weidong Mei, {\em Member, IEEE}, Changsheng You, \\{\em Member, IEEE}, Qingqing Wu, {\em Senior Member, IEEE}, and Jie Tang, {\em Senior Member, IEEE}} \vspace{-0.6cm}

\thanks{The work of Beixiong Zheng was supported in part by the National Natural Science Foundation of China under Grant 62571193 and Grant 62331022, the Guangdong program under Grant 2023QN10X446 and Grant 2023ZT10X148, and the GJYC program of Guangzhou under Grant 2024D01J0079 and Grant 2024D03J0006. \textit{(Corresponding author: Beixiong Zheng.)}}
\thanks{Qiyao Wang and Beixiong Zheng are with the School of Microelectronics, South China University of Technology, Guangzhou 511442, China (e-mails: 202420165372@mail.scut.edu.cn; bxzheng@scut.edu.cn).}
\thanks{Xue Xiong is with the School of Future Technology, South China University of Technology, Guangzhou 511442, China (e-mail: ftxuexiong@mail.scut.edu.cn).}
\thanks{Weidong Mei is with the National Key Laboratory of Science and Technology on Communications, University of Electronic Science and Technology of China, Chengdu 611731, China (e-mail: wmei@uestc.edu.cn).}
\thanks{Changsheng You is with the Department of Electronic and Electrical Engineering, Southern University of Science and Technology, Shenzhen 518055, China, and also with the Shenzhen Key Laboratory of Optoelectronics
and Intelligent Sensing, Shenzhen 518055, China (e-mail: youcs@sustech.edu.cn).}
\thanks{Qingqing Wu is with the Department of Electronic Engineering, Shanghai Jiao Tong University, Shanghai 200240, China (e-mail: qingqingwu@sjtu.edu.cn).}
\thanks{Jie Tang is with the School of Electronic and Information Engineering, South China University of Technology, Guangzhou 510640, China (e-mail: eejtang@scut.edu.cn).}
}}

\maketitle

\begin{abstract}
Enhancing communication reliability and computation efficiency remains a fundamental challenge for stringent low-latency mobile edge computing (MEC) services. Rotatable antenna (RA) is a promising technology that introduces new spatial degrees-of-freedom (DoFs) by independently adjusting antenna boresight directions to proactively improve wireless channel conditions for latency-critical users. In this letter, we investigate an RA-enabled MEC system and minimize the maximum computation latency by jointly optimizing the MEC server's computing resource allocation, receive beamforming, and the deflection angles of all RAs. To solve the resulting non-convex problem, we develop an efficient alternating optimization (AO) framework, where the edge computing resource allocation, receive beamforming, and RA deflection angles are respectively optimized via the Karush--Kuhn--Tucker (KKT) conditions, semidefinite relaxation (SDR) with bisection, and fractional programming (FP) with successive convex approximation (SCA). Simulation results verify that the proposed RA-enabled MEC scheme significantly reduces the maximum computation latency compared with conventional benchmarks.
\end{abstract}


\section{Introduction}
To meet the ambitious vision of the forthcoming sixth-generation (6G) network featuring ubiquitous connectivity and ultra-low latency, mobile edge computing (MEC) has been widely regarded as a promising technology that offloads computation tasks from resource-limited devices to nearby edge servers, thereby significantly reducing end-to-end latency and energy consumption \cite{8016573}, \cite{EnergyEfficient}. However, the performance of MEC systems is fundamentally constrained by the wireless links used for task offloading. The limited coverage, channel capacity constraints, and dynamic network conditions can severely degrade the quality of service (QoS), especially for latency-sensitive applications.

To overcome these challenges, rotatable antenna (RA) technology has recently emerged as an effective approach to enhance spatial degrees of freedom (DoFs) by enabling each antenna to flexibly adjust its three-dimensional (3D) boresight direction \cite{Conference-version, Modeling-and-Optimization, Rotatable-antenna-tutorial, Opportunities-and-Challenges}. RA offers a compact and low-cost solution suitable for edge deployments. Owing to these advantages, RA technology has demonstrated notable performance gains in various wireless communication scenarios, such as physical layer security \cite{Secure-Wireless}, integrated sensing and communication (ISAC) \cite{Integrated-Sensing}, and efficient channel estimation \cite{Efficient-Channel}. By flexibly adjusting antenna boresight directions, RAs can focus energy toward bottleneck users, mitigate unfavorable channel conditions, and improve offloading reliability in latency-sensitive MEC systems. Nevertheless, despite these promising advantages, the integration of RA technology into MEC architectures remains largely unexplored. In particular, how to jointly exploit RA directional control, wireless signal processing, and edge computing resource allocation to minimize computation latency remains an open problem.

Motivated by the above, we study an RA-enabled MEC uplink system, where an RA array is deployed at the base station (BS) to flexibly adjust antenna boresight directions toward different users. We formulate a max-latency minimization problem by jointly optimizing the MEC server's computing resource allocation, receive beamforming vectors, and RA deflection angles. To solve this non-convex problem, we develop an alternating optimization (AO) framework, where the computing resource allocation is derived via the Karush–Kuhn–Tucker (KKT) conditions, the receive beamforming is optimized by semidefinite relaxation (SDR) and bisection, and the RA deflection angles are updated using fractional programming (FP) and successive convex approximation (SCA) techniques. Simulation results verify that the proposed RA-enabled MEC scheme significantly reduces computation latency compared with conventional benchmarks.

\begin{figure}[t]
	\vspace{-0.2cm}
	\centering
	\includegraphics[width=2.75in,height=1.8in]{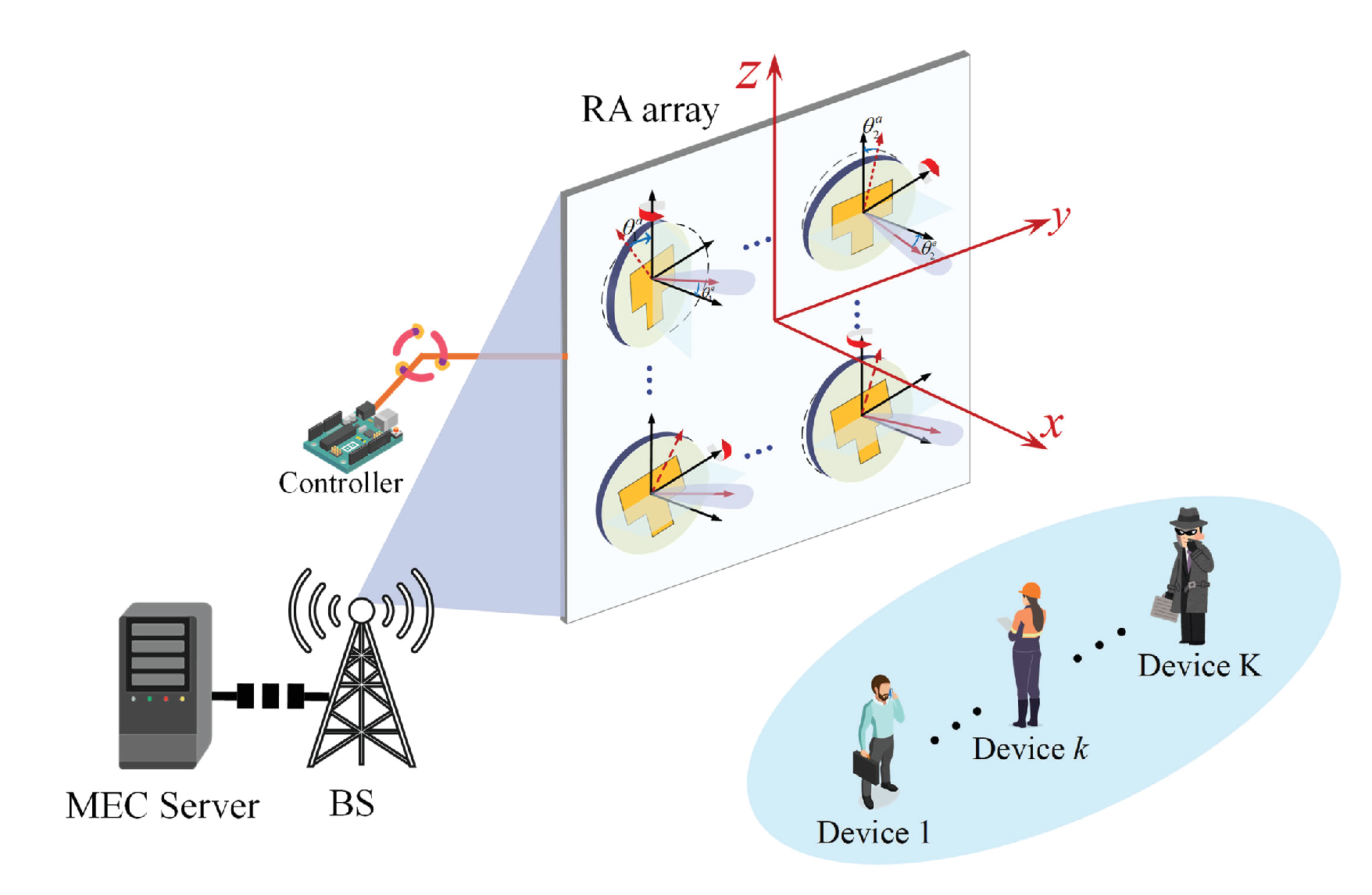}
	\vspace{-0.1cm}
     \captionsetup{singlelinecheck=false, justification=raggedright}
	\caption{An  RA-enabled MEC system.}
     \vspace{-0.1cm}
	\label{RA_MEC}  
	\vspace{-0.5cm}
\end{figure}

\section{System Model And Problem Formulation}

As shown in Fig.~\ref{RA_MEC}, we consider an RA-enabled MEC uplink communication system, where the BS equipped with an edge computing server serves $K$ devices. The devices simultaneously transmit data to the BS and may offload part or all of their computation tasks to the MEC server. The BS is equipped with a uniform planar array (UPA) of $N \triangleq N_yN_z$ directional RAs placed on the $y$-$z$ plane, where $N_y$ and $N_z$ denote the numbers of RAs along the $y$- and $z$-axes, respectively, while each device has a single isotropic antenna. The reference boresight of each RA is assumed to be parallel to the positive $x$-axis. Let $\mathbf{p}_n \in \mathbb{R}^{3 \times 1}$ and $\mathbf{q}_k \in \mathbb{R}^{3 \times 1}$ denote the reference positions of the $n$-th RA and the $k$-th device, respectively. The deflection angle vector of the $n$-th RA is defined as $\boldsymbol{\theta}_n \triangleq [\theta_n^{e}, \theta_n^{a}]^{T}$, $n = 1, \ldots, N$, where $\theta_n^{e}$ is the zenith angle between its boresight direction and the $x$-axis, and $\theta_n^{a}$ is the azimuth angle between the projection of its boresight vector onto the $y$-$z$ plane and the $z$-axis. Accordingly, the pointing vector of RA $n$ is defined as $\vec{\mathbf{f}}(\boldsymbol{\theta}_n) = [\cos(\theta_n^{e}), \sin(\theta_n^{e}) \sin(\theta_n^{a}), \sin(\theta_n^{e}) \cos(\theta_n^{a})]^{T}$. To account for the antenna boresight adjustment range, the zenith angle is constrained by $0 \leq \theta_n^{e} \leq \theta_{\text{max}}, \forall n$, where $\theta_{\text{max}}$ is the maximum allowable zenith angle. For notational convenience, we introduce $\vec{\mathbf{f}}_n \in \mathbb{R}^{3 \times 1}$ to represent the RA pointing vector associated with $\boldsymbol{\theta}_n$, i.e., $\vec{\mathbf{f}}_n = \vec{\mathbf{f}}(\boldsymbol{\theta}_n)$ with $\|\vec{\mathbf{f}}_n\| = 1$.

\vspace{-0.4cm}
\subsection{Communication Model}
\vspace{-0.1cm}
We consider an uplink communication model, where the computation offloading of the \textit{K} devices is performed over a predefined frequency bandwidth \textit{B} and shares the same time resource. All device-to-BS links are assumed to experience quasi-static flat-fading channels. Accordingly, the channel between device \textit{k} and RA \textit{n} can be modeled as

\vspace{-0.25cm}
\begin{equation}
\label{eq:definition}
h_{k,n}(\vec{\mathbf{f}}_n) = \sqrt{L(d_{k,n})} g_{k,n},
\end{equation}\normalsize

\vspace{-0.15cm}
\noindent
where $L(d_{k,n}) = \zeta_0 \left( d_0 / d_{k,n} \right)^{\alpha_0}$ represents the large-scale channel power gain, $\zeta_0$ is the channel
power gain at the reference distance of $d_0 = 1\ \text{meter (m)}$, $\alpha_0$ denotes the path loss exponent, and $d_{k,n}$ represents the distance from the $k$-th device to the $n$-th RA. 

Furthermore, we assume that $g_{k,n}$ follows an independent Rician fading model with Rician factor $\kappa_k$, given by

\vspace{-0.25cm}
\begin{equation}
	g_{k,n} = \sqrt{\frac{\kappa_k}{\kappa_k + 1}} \bar{g}_{k,n} +\sqrt{\frac{1}{\kappa_k + 1}} \tilde{g}_{k,n},
\end{equation}\normalsize

\vspace{-0.15cm}
\noindent
where $\bar{g}_{k,n} = \sqrt{G_{k,n}} e^{-j \frac{2\pi}{\lambda} d_{k,n}}$ represents the line-of-sight (LoS) channel component, $\tilde{g}_{k,n} \sim \mathcal{CN}(0, 1)$ denotes the non-LoS (NLoS) channel component characterized by Rayleigh fading, and $\lambda$ is the signal wavelength. In addition, $G_{k,n}$ denotes the generic directional radiation pattern for each RA, which can be modeled as \cite{Modeling-and-Optimization}

\vspace{-0.35cm}
\begin{equation}
\label{eq:G_model}
G(\epsilon, \phi) = 
\begin{cases} 
G_0 \cos^{2p}(\epsilon), & \epsilon \in [0, \pi/2),\ \phi \in [0,2\pi), \\ 0, & \text{otherwise,} 
\end{cases} 
\end{equation}\normalsize

\vspace{-0.15cm}
\noindent
where $\epsilon$ and $\phi$ are the zenith and azimuth angles of any spatial directions with respect to the RA's boresight direction, $G_0$ is the maximum gain in the boresight direction (i.e., $\epsilon = 0$) with $G_0 = 2(2p + 1)$ satisfying the law of power conservation, and $p \geq 0$ determines the directivity of antenna. Additionally, $\cos(\epsilon_{k,n}) \triangleq \vec{\mathbf{f}}_n^T \vec{\mathbf{q}}_{k,n}$ is the projection between the device $k$'s direction vector $\vec{\mathbf{q}}_{k,n} \triangleq \frac{\mathbf{q}_k - \mathbf{p}_n}{\|\mathbf{q}_k - \mathbf{p}_n\|}$ and the pointing vector of RA $n$. 

 Let $\mathbf{F} \triangleq \left[ \vec{\mathbf{f}}_1, \vec{\mathbf{f}}_2, \ldots, \vec{\mathbf{f}}_N \right] \in \mathbb{R}^{3 \times N}$ denote the RA deflection matrix. The overall channel from BS to device $k$ is given by
 
\vspace{-0.25cm}
\begin{equation}
\label{eq:hk_definition}
\mathbf{h}_k(\mathbf{F}) = \left[ h_{k,1}(\vec{\mathbf{f}}_1), h_{k,2}(\vec{\mathbf{f}}_2), \ldots, h_{k,N}(\vec{\mathbf{f}}_N) \right]^T.
\end{equation}\normalsize

\vspace{-0.15cm}
\noindent
Denote the uplink offloading transmit power and transmitted symbol of device $k$ by $P_k$ and $s_k$, respectively, the received signal at the BS is expressed as
\vspace{-0.1cm}
\begin{equation}
\mathbf{y} = \sum_{k=1}^{K} \mathbf{h}_k(\mathbf{F}) \sqrt{P_k} s_k + \mathbf{n},
\end{equation}\normalsize

\vspace{-0.1cm}
\noindent
where $\mathbf{n} \sim \mathcal{CN}(0, \sigma^2 \mathbf{I}_N)$ represents the additive white Gaussian  noise (AWGN) with zero-mean and variance $\sigma^2$. Upon receiving $\mathbf{y}$, the BS applies a linear receive beamforming vector $\mathbf{w}_k^H \in \mathbb{C}^{1 \times N}$ with $\|\mathbf{w}_k\| = 1$ to extract the signal of device $k$, i.e.,

\vspace{-0.45cm}
\begin{equation}
y_k \!=\! \mathbf{w}_k^H \mathbf{h}_k(\mathbf{F}) \sqrt{P_k} s_k \!+\! \sum_{\substack{j=1,j \neq k}}^{K} \mathbf{w}_k^H \mathbf{h}_j(\mathbf{F}) \sqrt{P_j} s_j \!+\! \mathbf{w}_k^H \mathbf{n}.
\end{equation}\normalsize

\vspace{-0.1cm}
Then, the signal-to-interference-plus-noise ratio (SINR) for decoding the signal from the $k$-th device is given by
\vspace{-0.1cm}
\begin{equation}
\label{eq:SINR}
\gamma_k(\mathbf{w}_k, \mathbf{F}) = \frac{{P}_k \left| \mathbf{w}_k^H \mathbf{h}_k(\mathbf{F}) \right|^2}{\sum_{\substack{j=1, j \neq k}}^{K} {P}_j \left| \mathbf{w}_k^H \mathbf{h}_j(\mathbf{F}) \right|^2 + \sigma^2}.
\end{equation}\normalsize

\vspace{-0.1cm}
\noindent
Accordingly, the maximum achievable computation offloading rate of the $k$-th device is given by

\vspace{-0.3cm}
\begin{equation}
R_k(\mathbf{w}_k, \mathbf{F}) = B\log_2\left(1 + \gamma_k(\mathbf{w}_k, \mathbf{F})\right).
\end{equation}\normalsize

\vspace{-0.3cm}
\subsection{Computing Model}
Each device $k$ has a computation task of $L_k$ bits, with each bit requiring $c_k$ CPU cycles. The task can be divided into two parts: one computed locally and the other offloaded to the edge server.

\indent \textit{1) Local Computing:} Let $\ell_k$ and $f_k^l$ denote the offloaded task size and CPU frequency of device $k$, respectively. The local computation time is given by
$D_k^l(\ell_k) = (L_k - \ell_k)c_k/f_k^l.$

\noindent
\indent \textit{2) Edge Computing:} We assume that edge computing for device $k$ starts after all its task bits have been offloaded. The edge latency consists of the transmission time from the device to the BS and the computation time at the MEC server. The MEC server allocates computing resource $f_k^e$ to device $k$, satisfying
\vspace{-0.1cm}
\begin{equation}
	\sum_{k=1}^{K} f_k^e \leq F_{\text{max}}, \quad f_k^e \geq 0, \ \forall k,
\end{equation}
\noindent
where $F_{\text{max}}$ denotes the maximum computing capacity of the MEC server. Since the computation result is assumed to be sufficiently small, the feedback delay is neglected. Thus, the edge latency is
$D_k^e(\mathbf{w}_k, \mathbf{F}, \ell_k, f_k^e) = \ell_k/R_k(\mathbf{w}_k, \mathbf{F}) + \ell_k c_k/f_k^e.$

Since local computation and task offloading can be performed simultaneously, the task completion time is determined by the slower branch. Hence, the computation latency of device $k$ is
$D_k(\mathbf{w}_k, \mathbf{F}, \ell_k, f_k^e) = \max\{D_k^l(\ell_k), D_k^e(\mathbf{w}_k, \mathbf{F}, \ell_k, f_k^e)\}$,
and the maximum computation latency is
$\tau^{\text{comp}} = \max\{D_k(\mathbf{w}_k, \mathbf{F}, \ell_k, f_k^e), \forall k\}$.


\vspace{-0.2cm}
\subsection{Problem Formulation}
In this paper, we aim to minimize the maximum computation latency $\tau^{\text{comp}}$ by jointly optimizing the offloaded data size $\boldsymbol{\ell} = [\ell_1, \ell_2, \ldots, \ell_K]^T$, edge computing resource allocation $\boldsymbol{f}^e = [f_1^e, f_2^e, \ldots, f_K^e]^T$, BS receive beamforming matrix $\mathbf{W} = [\mathbf{w}_1, \mathbf{w}_2, \ldots, \mathbf{w}_K]^T$, and RA pointing matrix $\mathbf{F}$. The associated optimization problem is formulated as

\vspace{-0.4cm}
{\small
	\begin{subequations} \label{prob:P1_main}
		\begin{align}
			\text{(P1)} \quad 
			&\min_{\boldsymbol{\ell}, \boldsymbol{f}^e, \mathbf{W}, \mathbf{F}} 
			\ \tau^{\text{comp}} \label{eq:P1_obj} \\
			&\text{s.t.} \ ||\mathbf{w}_k|| = 1, \ \forall k, \label{eq:P1_const1}\\ 
			&\quad 0 \leq \arccos(\vec{\mathbf{f}}_n^T \mathbf{e}_1) 
			\leq \theta_{\text{max}}, \ \forall n, \label{eq:P1_const2}\\ 
			&\quad \|\vec{\mathbf{f}}_n\| = 1, \ \forall n, \label{eq:P1_const3} \\ 
			&\quad \ell_k \in \{0, 1, \ldots, L_k\}, \quad k = 1, 2, \ldots, K, 
			\label{eq:P1_const4}\\ 
			&\quad \sum_{k=1}^{K} f_k^e \leq F_{\text{max}}, \label{eq:P1_const5} \\ 
			&\quad f_k^e \geq 0, \quad k = 1, 2, \ldots, K. \label{eq:P1_const6} 
		\end{align}
	\end{subequations}
}

\vspace{-0.1cm}
\noindent
where \eqref{eq:P1_const1} imposes the unit-norm constraints on receive beamforming vectors; \eqref{eq:P1_const2} and \eqref{eq:P1_const3} constrain the RA deflection range and unit-norm pointing vectors, respectively; \eqref{eq:P1_const4} ensures that the offloaded data size is an integer within $[0,L_k]$; and \eqref{eq:P1_const5}--\eqref{eq:P1_const6} restrict the edge computing resources. Due to the strongly coupled variables and non-convex constraints, problem (P1) is difficult to solve directly. Thus, we propose an AO algorithm to iteratively optimize the four variable blocks.

\section{Proposed Algorithm For Problem (P1)}
\vspace{-0.05cm}
In this section, we develop an AO-based algorithm for solving problem (P1), which alternately optimizes the computation variables $\{\boldsymbol{\ell}, \boldsymbol{f}^e\}$ and communication variables $\{\mathbf{W}, \mathbf{F}\}$ until convergence. Accordingly, problem (P1) is decomposed into two tractable subproblems.

\vspace{-0.2cm}
\subsection {Joint Optimization of Offloaded Data Size and Edge Computing Resource Allocation}
\label{sec:subproblem1}

Given the receive beamforming matrix $\mathbf{W}$ and the RAs' pointing matrix $\mathbf{F}$, problem (P1) reduces to
\vspace{-0.1cm}
\begin{align}
	\text{(P2)} \quad &\min_{\boldsymbol{\ell}, \boldsymbol{f}^e} \ \tau^{\text{comp}} \label{eq:P1_1_obj} \\
	&\text{s.t.} \ \eqref{eq:P1_const4}, \eqref{eq:P1_const5}, \eqref{eq:P1_const6}. \notag
\end{align}

\vspace{-0.1cm}
\noindent
Although simplified, problem (P2) remains difficult to solve optimally due to the coupling between $\boldsymbol{\ell}$ and $\boldsymbol{f}^e$ in \eqref{eq:P1_const4}--\eqref{eq:P1_const6}. To decouple them, we first fix $\boldsymbol{f}^e$ and optimize $\boldsymbol{\ell}$, where (P2) reduces to $K$ single-variable optimization problems with respect to $\ell_k$. Each admits a closed-form solution \cite{bai2020latency}, given by
\vspace{-0.3cm}
\begin{equation}
	\label{eq:integerize}
	\ell_k^{\star} = \operatorname*{argmin}_{\hat{\ell}_k \in \left\{ \lfloor \hat{\ell}_k^{\star} \rfloor, \; \lceil \hat{\ell}_k^{\star} \rceil \right\}} D_k(\hat{\ell}_k),
\end{equation}

\vspace{-0.15cm}
\noindent
where $\lfloor\cdot\rfloor$ and $\lceil\cdot\rceil$ denote the floor and ceiling operations, respectively, and
\vspace{-0.15cm}
\begin{equation}
	\label{eq:computing_l_hat_star}
	\hat{\ell}_k^{\star} = \frac{L_k c_k R_k f_k^e}{f_k^e f_k^l + c_k R_k (f_k^e + f_k^l)}. 
\end{equation}

\noindent
By substituting $\hat{\ell}_k^{\star}$ into \eqref{eq:P1_obj} and introducing the slack variable $t_1$, problem (P2) is transformed into
{\small
	\begin{subequations} \label{prob:P2_main}
		\begin{align}
			\text{(P2.1)} \quad &\min_{\boldsymbol{f}^e, t_1} \ t_1 \label{eq:P2_obj} \\
			&\text{s.t.} \ \frac{ L_k c_k^2 R_k + L_k c_k f_k^e }{f_k^e f_k^l + c_k R_k \left( f_k^e + f_k^l \right)} \leq t_1, \ \forall k, \label{eq:P2_const1} \\
			&\quad \ \eqref{eq:P1_const5}, \eqref{eq:P1_const6}. \notag
		\end{align}
	\end{subequations}
}
\vspace{-0.4cm}

It can be verified that problem (P2.1) is convex, since the left-hand side of \eqref{eq:P2_const1} is convex with respect to $f_k^e$, and the other constraints are affine. Thus, the KKT conditions can be used to derive the closed-form solution of $f_k^e$. Due to the min--max latency objective, only the multiplier associated with the bottleneck device is active, while the others are zero. Without loss of generality, the active multiplier is normalized to unity. Accordingly, for a given dual variable $\mu$, the optimal $f_k^e$ is given in closed form as in \cite{bai2020latency}
\vspace{-0.2cm}
{\small
	\begin{equation}
		\label{eq:optimal_fke}
		f_k^e = \frac{\sqrt{\frac{L_k c_k^3 R_k^2}{\mu}} - c_k R_k f_k^l}{f_k^l + c_k R_k}, \quad k = 1, \ldots, K,
	\end{equation}
}
\vspace{-0.1cm}
\noindent
where $\mu^{\star}$ denotes the optimal dual variable, which can be found via bisection over $(\mu_l, \mu_u] = \left(0, \min\limits_{k} \left( \frac{L_k c_k}{{f_k^l}^2} \right) \right]$.

\vspace{-0.2cm}
\subsection{Joint Optimization of Receive Beamforming and Pointing Matrix}
Given the offloaded data size $\boldsymbol{\ell}$ and edge computing resource allocation $\boldsymbol{f}^e$, problem (P1) is reformulated for optimizing the receive beamforming matrix $\mathbf{W}$ and RA pointing matrix $\mathbf{F}$ as
\vspace{-0.45cm}
\begin{align}
	\text{(P3)} \quad &\min_{\mathbf{W}, \mathbf{F}} \ \tau^{\text{comp}} \label{eq:P3_obj} \\
	&\text{s.t.} \ \eqref{eq:P1_const1}, \eqref{eq:P1_const2}, \eqref{eq:P1_const3}. \notag
\end{align}

\vspace{-0.1cm}
As shown in Section~\ref{sec:subproblem1}, at the optimal solution to problem (P2), $D_k^l = D_k^e$, and thus the objective in \eqref{eq:P3_obj} can be equivalently expressed by the edge computation latency. By replacing $D_k$ with $D_k^e$ and $\ell_k$ with the closed-form $\hat{\ell}_k^{\star}$ in \eqref{eq:computing_l_hat_star}, problem (P3) reduces to
\vspace{-0.1cm}
{\small
	\begin{align}
		\text{(P3.1)} \quad 
		&\min_{\mathbf{W}, \mathbf{F}} 
		\max_{k} \ 
		\frac{L_k c_k f_k^e}
		{f_k^e f_k^l + c_k R_k(\mathbf{w}_k, \mathbf{F}) (f_k^e + f_k^l)}
		+ \frac{\ell_k c_k}{f_k^e} \label{eq:P4_obj}\\
		&\text{s.t.} \ \eqref{eq:P1_const1}, \eqref{eq:P1_const2}, \eqref{eq:P1_const3}. \notag
	\end{align}
}

Moreover, to handle the fractional SINR expression in \eqref{eq:SINR}, we introduce slack variables $t_2$ and $\{t_{3,k}\}$, and reformulate problem (P3.1) as
\vspace{-0.1cm}
{\small
	\begin{subequations} \label{prob:P5_main}
		\begin{align}
			\text{(P3.2)} \quad 
			&\min_{\mathbf{W}, \mathbf{F}, t_2, \{t_{3,k}\}} \ t_2 \label{eq:P5_obj}\\
			&\text{s.t.} \ \frac{\ell_k c_k}{f_k^e} \leq t_{3,k}, \ \forall k, \label{eq:P5_c1} \\
			&\quad \frac{L_k c_k f_k^e}{f_k^e f_k^l + c_k R_k(\mathbf{w}_k, \mathbf{F}) (f_k^e + f_k^l)}
			+ t_{3,k} \leq t_2, \ \forall k, \label{eq:P5_c2} \\
			&\quad R_k \leq B\log_2(1 + \gamma_k), \ \forall k, \label{eq:P5_c3} \\
			&\quad \gamma_k \leq 
			\frac{{P}_k \left| \mathbf{w}_k^H \mathbf{h}_k(\mathbf{F}) \right|^2}
			{\sum_{j\neq k}^{K} {P}_j \left| \mathbf{w}_k^H \mathbf{h}_j(\mathbf{F}) \right|^2 + \sigma^2}, \ \forall k, 
			\label{eq:P5_c4} \\
			&\quad \eqref{eq:P1_const1}, \eqref{eq:P1_const2}, \eqref{eq:P1_const3}. \notag
		\end{align}
	\end{subequations}
}

\vspace{-0.1cm}
\noindent
Problem (P3.2) remains non-convex due to variable coupling and the fractional SINR constraint in \eqref{eq:P5_c4}. We thus decompose it into two subproblems.

\noindent 1) \textit{Receive Beamforming Optimization}

Given the pointing matrix $\mathbf{F}$, the SINR constraints reduce to
\vspace{-0.1cm}
{\small
	\begin{equation}
		\gamma_k \leq 
		\frac{{P}_k \left| \mathbf{w}_k^H \mathbf{h}_k \right|^2}
		{\sum_{j\neq k}^{K} {P}_j \left| \mathbf{w}_k^H \mathbf{h}_j \right|^2 + \sigma^2}, \ \forall k.
	\end{equation}
}

\vspace{-0.1cm}
\noindent
To facilitate convex reformulation, define $\mathbf{W}_k = \mathbf{w}_k \mathbf{w}_k^H$, which satisfies $\mathbf{W}_k \succeq \boldsymbol{0}$ and $\operatorname{rank}(\mathbf{W}_k)=1$. Then, $\left| \mathbf{w}_k^H \mathbf{h}_k \right|^2 = \operatorname{tr}(\mathbf{W}_k \mathbf{H}_{k})$ and $\left| \mathbf{w}_k^H \mathbf{h}_j \right|^2 = \operatorname{tr}(\mathbf{W}_k \mathbf{H}_{j})$, where $\mathbf{H}_{k} = \mathbf{h}_k \mathbf{h}_k^H$ and $\mathbf{H}_{j} = \mathbf{h}_j \mathbf{h}_j^H$.

By relaxing the non-convex rank-one constraint, we obtain the following SDR problem:
\vspace{-0.1cm}
{\small
	\begin{subequations} \label{prob:P7_main}
		\begin{flalign}
			\text{(P3.3)} \quad &\min_{\mathbf{W}, t_2, \{t_{3,k}\}} \ t_2 \label{eq:P7_obj}\\
			\text{s.t.} &\ \gamma_k \left(\sum_{j\neq k}^{K} P_j \operatorname{tr}(\mathbf{W}_k \mathbf{H}_{j}) + \sigma^2\right) 
			\leq P_k \operatorname{tr}(\mathbf{W}_k \mathbf{H}_{k}), \forall k, \label{eq:P7_c1}\\
			&\operatorname{tr}(\mathbf{W}_k) = 1,\quad \forall k, \label{eq:P7_c2}\\
			&\mathbf{W}_k \succeq \boldsymbol{0}, \forall k, \label{eq:P7_c3}\\
			&\eqref{eq:P5_c1}, \eqref{eq:P5_c2}, \eqref{eq:P5_c3}. \notag
		\end{flalign}
	\end{subequations}
}

\vspace{-0.1cm}
\noindent
Problem (P3.3) is quasi-convex with respect to $\{\mathbf{W}_k\}$ and can be solved via bisection search. Specifically, for any given $\gamma_k$, the associated feasibility problem is a convex SDP. Although SDR may yield a high-rank solution, a high-quality rank-one approximation can be recovered by Gaussian randomization to obtain the receive beamforming vectors.

\vspace{0.05cm}
\noindent 2) \textit{Pointing Matrix Optimization}

For a given receive beamforming matrix $\mathbf{W}$, problem (P3.2) is equivalently expressed as
\vspace{-0.1cm}
{\small
	\begin{subequations} \label{prob:P8_main}
		\begin{align}
			\text{(P3.4)} \quad 
			&\min_{\mathbf{F}, t_2, \{t_{3,k}\}} \ t_2 \label{eq:P8_obj}\\
			\text{s.t.} \quad
			&\gamma_k \leq 
			\frac{{P}_k \left| \mathbf{w}_k^H \mathbf{h}_k(\mathbf{F}) \right|^2}
			{\sum_{j=1,j\neq k}^{K} {P}_j \left| \mathbf{w}_k^H \mathbf{h}_j(\mathbf{F}) \right|^2 + \sigma^2}, \ \forall k, \label{eq:P8_c1}\\
			&\cos(\theta_{\max}) \leq \vec{\mathbf{f}}_n^T \mathbf{e}_1 \leq 1, \ \forall n, \label{eq:P8_c2}\\
			&\eqref{eq:P1_const3}, \eqref{eq:P5_c1}, \eqref{eq:P5_c2}, \eqref{eq:P5_c3}. \notag
		\end{align}
	\end{subequations}
}

\vspace{-0.1cm}
\noindent
where \eqref{eq:P8_c2} is equivalent to \eqref{eq:P1_const2} and restricts RA's zenith angle within $[0,\theta_{\max}]$.

Due to the fractional SINR constraint in \eqref{eq:P8_c1}, we apply the quadratic transform, yielding
\vspace{-0.1cm}
{\small
	\begin{equation}
		\begin{aligned}
			\label{eq:quadratic transform}
			f_k(\mathbf{F}, \eta_k)
			= \ &2 \operatorname{\mathcal{R}} \left\{ \eta_k \sqrt{P_k} \mathbf{w}_k^H \mathbf{h}_k(\mathbf{F}) \right\} \\
			&- |\eta_k|^2 \left( \sum_{j=1,j\neq k}^{K} {P}_j 
			\left| \mathbf{w}_k^H \mathbf{h}_j(\mathbf{F}) \right|^2 + \sigma^2 \right),
		\end{aligned}
	\end{equation}
}

\vspace{-0.1cm}
\noindent
where $\eta_k$ is the auxiliary variable introduced by the quadratic transform for device $k$. We then iteratively optimize $\mathbf{F}$ and $\{\eta_k\}$ to minimize \eqref{eq:P8_obj}. For a given $\mathbf{F}$, the optimal $\eta_k$ is given by \cite{Fractional-Programming}
\vspace{-0.2cm}
\begin{equation}
	\eta_k^{\star} = 
	\frac{ \sqrt{P_k} \mathbf{w}_k^H \mathbf{h}_k(\mathbf{F})}
	{\sum_{j=1,j\neq k}^{K} {P}_j \left| \mathbf{w}_k^H \mathbf{h}_j(\mathbf{F}) \right|^2 + \sigma^2}.
\end{equation}

\vspace{-0.1cm}
\noindent
For fixed $\eta_k^{\star}$, the remaining problem with respect to $\mathbf{F}$ is still non-convex due to the nonlinear dependence of $\mathbf{h}_k(\mathbf{F})$ on the RA pointing vectors. We thus adopt SCA to construct a convex approximation of problem (P3.4). Let $\mathbf{F}^{(i)}$ denote the RA pointing matrix at the $i$-th iteration. By the first-order Taylor expansion at $\mathbf{F}^{(i)}$, the term $\mathbf{w}_k^H \mathbf{h}_k(\mathbf{F})$ in \eqref{eq:quadratic transform} is approximated as
\vspace{-0.3cm}
\begin{equation}
	\begin{aligned}
		\Phi_k^{(i+1)}(\mathbf{F}) \triangleq 
		\mathbf{w}_k^H \mathbf{h}_k(\mathbf{F}^{(i)})
		+\sum_{n=1}^{N} w_{k,n}^* (\mathbf{h}_{k,n}^{\prime})^T
		(\vec{\mathbf{f}}_n-\vec{\mathbf{f}}_n^{(i)}),
	\end{aligned}
\end{equation}

\vspace{-0.15cm}
\noindent
where $\mathbf{h}_{k,n}^{\prime} = \frac{\partial h_{k,n}(\vec{\mathbf{f}}_n^{(i)})}{\partial \vec{\mathbf{f}}_n^{(i)}} = \left( \tilde{\beta}_{k,n} p \left( (\vec{\mathbf{f}}_n^{(i)})^T \vec{\mathbf{q}}_{k,n} \right)^{p-1} \right) \vec{\mathbf{q}}_{k,n}$, with $\tilde{\beta}_{k,n} \triangleq 
\sqrt{\frac{\kappa_k}{\kappa_k + 1}}
\sqrt{L(d_{k,n})G_0}\,
e^{-j \frac{2\pi}{\lambda} d_{k,n}}$, and $(\cdot)^*$ denotes conjugation.

Furthermore, to handle the non-convex constraint in \eqref{eq:P8_c1}, we construct a quadratic convex upper bound for $u_{k,j}(\mathbf{F}) \triangleq \left| \mathbf{w}_k^H \mathbf{h}_j(\mathbf{F}) \right|^2$. Let $\nabla u_{k,j}(\vec{\mathbf{f}}_n)$ and $\nabla^2 u_{k,j}(\vec{\mathbf{f}}_n)$ denote the gradient and Hessian of $u_{k,j}$ with respect to $\vec{\mathbf{f}}_n$, respectively, as derived in Appendix~\ref{app:Derivations}. For $\delta_{k,j,n}>0$ satisfying $\delta_{k,j,n}\mathbf{I}_3 \succeq \nabla^2 u_{k,j}(\vec{\mathbf{f}}_n)$, whose closed-form expression is given in Appendix~\ref{app:Construction}, the following quadratic upper bound holds:
\vspace{-0.1cm}
\begin{equation}
	\begin{aligned}
		u_{k,j}(\vec{\mathbf{f}}_n) 
		&\le u_{k,j}(\vec{\mathbf{f}}_n^{(i)}) 
		+ \nabla u_{k,j}(\vec{\mathbf{f}}_n^{(i)})^T 
		(\vec{\mathbf{f}}_n - \vec{\mathbf{f}}_n^{(i)}) \\
		&\quad + \frac{\delta_{k,j,n}}{2}
		(\vec{\mathbf{f}}_n - \vec{\mathbf{f}}_n^{(i)})^T
		(\vec{\mathbf{f}}_n - \vec{\mathbf{f}}_n^{(i)}) \\
		&\triangleq u_{k,j}^{\text{ub}}(\vec{\mathbf{f}}_n), 
		\label{eq:upper_bound}
	\end{aligned}
\end{equation}

\vspace{-0.05cm}
\noindent
which follows from \cite[Lemma 12]{sun2016majorization}. Then, \eqref{eq:quadratic transform} is approximated as
\vspace{-0.1cm}
\begin{equation}
	\begin{aligned}
		f_k(\mathbf{F}, \eta_k) \geq \ 
		&2 \operatorname{\mathcal{R}} \left\{ \eta_k^H \sqrt{P_k} \Phi_k^{(i+1)}(\mathbf{F}) \right\} \\
		&- |\eta_k|^2 \left( \sum_{j=1,j\neq k}^{K} P_j u_{k,j}^{\text{ub}}(\mathbf{F}) + \sigma^2 \right),
	\end{aligned}
	\vspace{-0.15cm}
\end{equation}

\vspace{-0.05cm}
\noindent
where $u_{k,j}^{\text{ub}}(\mathbf{F})$ denotes the constructed convex upper bound applied to the RA pointing matrix $\mathbf{F}$.

\subsection{Overall Algorithm}
\vspace{-0.2cm}
\begin{algorithm}
\caption{Joint Optimization of $\boldsymbol{\ell}, \boldsymbol{f}^e, \mathbf{W}, \mathbf{F}$.}
\footnotesize 
\label{alg:proposed_ao}
\setstretch{0.9}
\begin{algorithmic}
\renewcommand{\algorithmicrequire}{\textbf{Initialize}}\REQUIRE
	\STATE Set $\mathbf{F}^{(0)} \!=\! [\mathbf{e}_1, \dots, \mathbf{e}_1]_{3\times N}$, threshold $\varepsilon > 0$, and $i \!=\! 0$.
	\STATE Compute $\boldsymbol{\ell}^{(0)}$ and $\boldsymbol{f}^{e(0)}$ by the closed-form solutions in (\ref{eq:computing_l_hat_star}) and (\ref{eq:optimal_fke}).
    \REPEAT
    \STATE Compute $\mathbf{W}^{(i+1)}$ by solving problem (P3.3) with given $\boldsymbol{\ell}^{(i)}$, $\boldsymbol{f}^{e(i)}$, and $\mathbf{F}^{(i)}$.
    \STATE Update $\mathbf{F}^{(i+1)}$ by solving problem (P3.4) with given $\boldsymbol{\ell}^{(i)}$, $\boldsymbol{f}^{e(i)}$, and $\mathbf{W}^{(i+1)}$.
	\STATE Calculate $\boldsymbol{\ell}^{(i+1)}$ and $\boldsymbol{f}^{e(i+1)}$ by the closed-form solution in (\ref{eq:computing_l_hat_star}) and (\ref{eq:optimal_fke}) with given $\mathbf{W}^{(i+1)}$ and $\mathbf{F}^{(i+1)}$.
	\STATE Update $i = i + 1$.
	\UNTIL $\left| \frac{\tau^{\text{comp}(i+1)} - \tau^{\text{comp}(i)}}{\tau^{\text{comp}(i)}} \right| \le \varepsilon.$
	\STATE \quad Integerize $\ell^{(i)}$ by (\ref{eq:integerize}).
\renewcommand{\algorithmicrequire}{\textbf{Output}}\REQUIRE $\boldsymbol{\ell}^{(i)}$, $\boldsymbol{f}^{e(i)}$, $\mathbf{W}^{(i)}$ and $\mathbf{F}^{(i)}$.
\end{algorithmic}
\end{algorithm}
The overall AO algorithm for solving (P1) is summarized in Algorithm~\ref{alg:proposed_ao}. Since $\tau^{\text{comp}}$ is non-increasing over iterations and lower bounded, the proposed AO algorithm is guaranteed to converge. In each iteration, the complexity of solving (P2) is $\mathcal{O}\!\left(I \log_2\!\left(\frac{\mu_u-\mu_l}{\epsilon_1}\right)K\right)$, where $\epsilon_1$ is the bisection accuracy and $I$ is the number of inner iterations. The complexities of solving (P3.3) via SDR and bisection and (P3.4) via SCA are $\mathcal{O}(N^{4.5}\ln(1/\epsilon_2))$ and $\mathcal{O}(N^{3.5}\ln(1/\epsilon_3))$, respectively, where $\epsilon_2$ and $\epsilon_3$ are the corresponding accuracy thresholds. Therefore, the overall complexity of solving (P1) is $\mathcal{O}\!\left(J\left(I\log_2\!\left(\frac{\mu_u-\mu_l}{\epsilon_1}\right)K+N^{4.5}\ln(1/\epsilon_2)+N^{3.5}\ln(1/\epsilon_3)\right)\right)$, where $J$ denotes the number of outer AO iterations.

\vspace{-0.1cm}
\section{Simulation Results}
In this section, we present simulation results to verify the proposed RA-enabled MEC uplink system with the proposed AO algorithm. The RA array is centered at the origin with inter-antenna spacing $d=\lambda/2$, $p=4$, and $\theta_{\max}=\pi/6$. The system operates at $2.4$ GHz ($\lambda=0.125\ \text{m}$) with noise power $\sigma^2=-60\ \text{dBm}$. Unless otherwise specified, we set $N=9$, $\zeta_0=-30\ \text{dB}$, $\alpha_0=2.8$, $\kappa_k=1,\forall k$, $B=2\ \text{MHz}$, $L_k=1000\ \text{Kb}$, $c_k=1000\ \text{cycle/bit}$, and $f_k^l=6\times10^8\ \text{cycle/s}$ for all devices. The offloading powers are identical, i.e., $P_k=P,\forall k$, and the devices are uniformly distributed over a semicircle centered at the origin with radius $r=40\ \text{m}$.

For comprehensive evaluation, the proposed RA-enabled MEC system is referred to as the \textbf{RA-enabled scheme}, and three benchmarks are considered: 1) \textbf{Fixed directional antenna scheme:} all RA deflections are fixed as $\vec{\mathbf{f}}_n=\mathbf{e}_1$, $\forall n$, where $\mathbf{e}_1=[1,0,0]^T$; 2) \textbf{Isotropic antenna scheme:} the directional gain is set to $G_0=1$ with $p=0$ in \eqref{eq:G_model}; and 3) \textbf{Random-orientation scheme:} each RA orientation is randomly generated within the rotational range specified in \eqref{eq:P1_const2}.

\begin{figure*}[!tbp]
    \vspace{-0.5cm}
    \centering
    \begin{minipage}[t]{0.315\textwidth}
        \centering
        \includegraphics[width=\linewidth,height=1.8in]{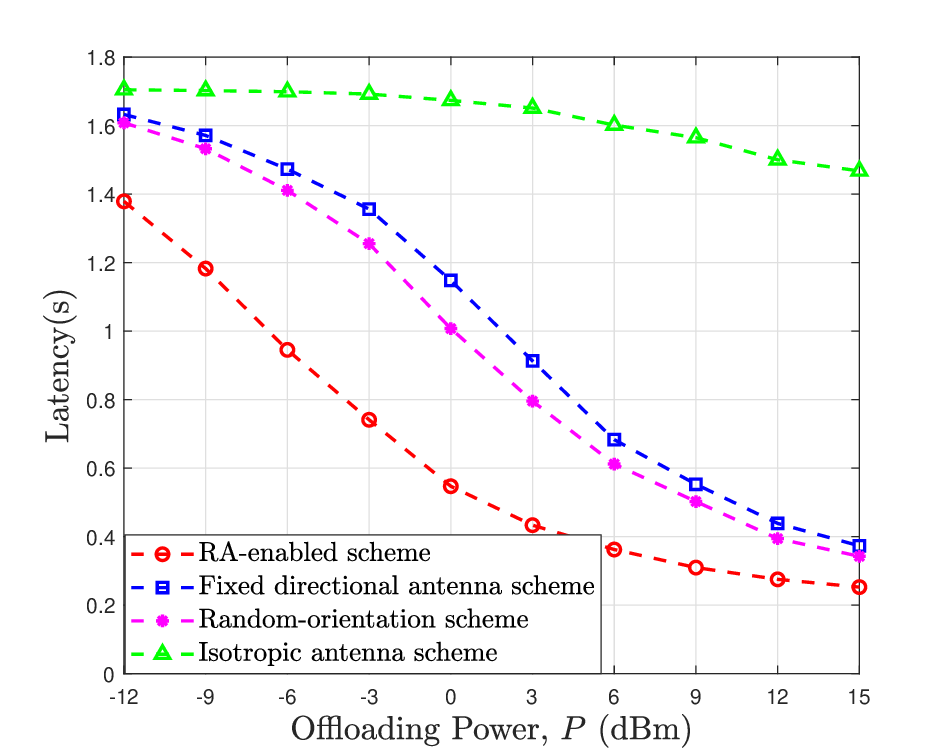}
        \caption{The maximum computation latency versus offloading power.}
        \label{fig.2}
    \end{minipage}%
    \vspace{-0.2cm}
    \hspace{3mm}
    \begin{minipage}[t]{0.315\textwidth}
        \centering
        \includegraphics[width=\linewidth,height=1.8in]{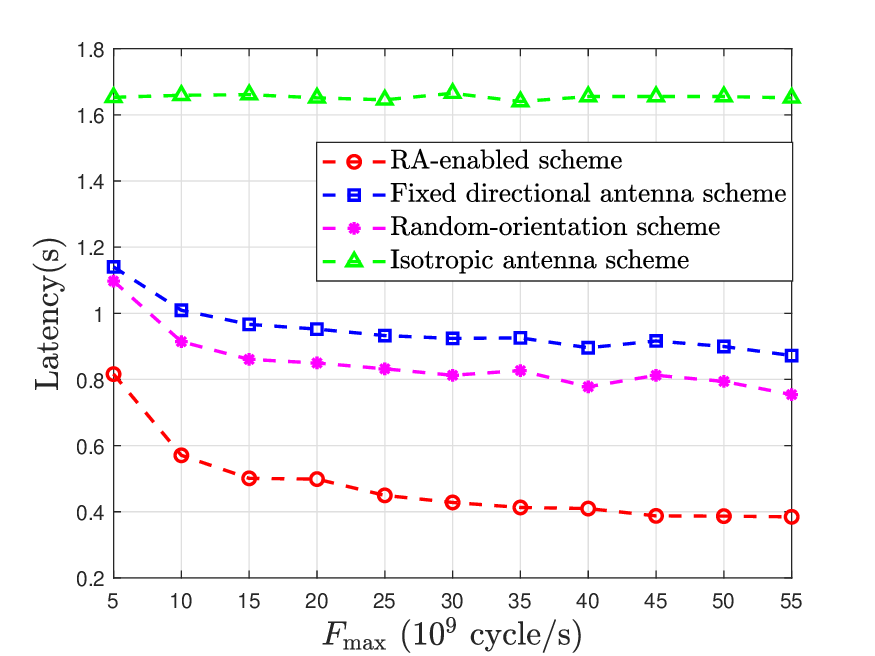}
        \caption{The maximum computation latency versus the maximum computing capacity.}
        \label{fig.3}
    \end{minipage}%
    \hspace{3mm}
    \begin{minipage}[t]{0.315\textwidth}
        \centering
        \includegraphics[width=\linewidth,height=1.8in]{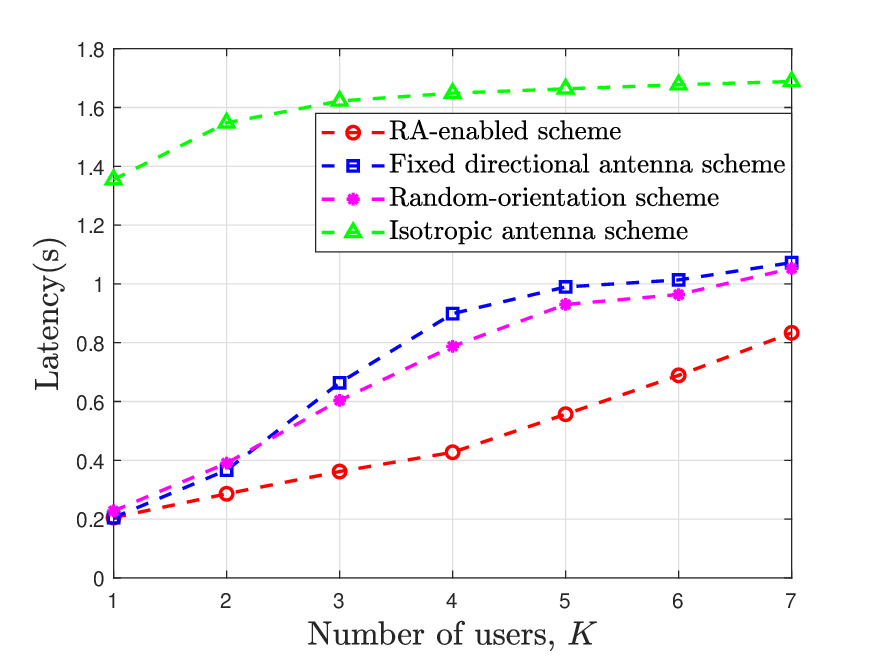}
        \caption{The maximum computation latency versus number of devices.}
        \label{fig.4}
    \end{minipage}
    \vspace{-0.2cm}
\end{figure*}

In Fig.~\ref{fig.2}, we show the latencies of different schemes versus the offloading power $P$ for $K = 4$ and $F_{\text{max}} = 30\times 10^9\ \text{cycle/s}$. It is observed that the latencies of all schemes decrease with increasing $P$, since a higher power improves the uplink offloading rate. The performance gap between the proposed RA-enabled scheme and the benchmark schemes becomes less pronounced at very low and high offloading power. At low power, noise dominates the link quality, making the directional gain insufficient to produce noticeable improvement. At high power, the latency is mainly constrained by the maximum computing capacity, and the offloading link is no longer the bottleneck, making the array directional gain provided by RA less critical. Furthermore, the RA-enabled scheme always outperforms the benchmark schemes, primarily owing to its ability to adaptively rotate the boresight directions, enhance the effective uplink channel gain, and increase the offloading rate. In contrast, the isotropic and fixed directional antenna schemes exhibit higher latency due to the absence of directional enhancement, while the random-orientation scheme fails to fully exploit the directional gain owing to its non-optimized antenna orientations.

Fig.~\ref{fig.3} illustrates the maximum computation latency versus the maximum computing capacity $F_{\text{max}}$ for $K = 4$ and $P = 3\ \text{dBm}$. As $F_{\text{max}}$ increases, the latencies of all schemes decrease due to the shortened task execution time at the edge server. When $F_{\text{max}}$ is limited, computing capability is the bottleneck of latency reduction. When $F_{\text{max}}$ is sufficiently high, the latency reduction becomes marginal, indicating that the offloading link plays a dominant role. Therefore, it is unnecessary to equip the edge computing node with an extremely powerful computing capability for latency minimization.

Fig.~\ref{fig.4} shows the latencies of different schemes versus the number of devices $K$ for $P = 3\ \text{dBm}$ and $F_{\text{max}} = 30\times 10^9\ \text{cycle/s}$. It is observed that the RA-enabled scheme consistently outperforms all benchmark schemes, while the latency of each scheme increases with $K$. This is partially due to the reduced edge computational resources allocated to each device. Moreover, antenna deflection adjustment becomes more challenging, making it difficult to simultaneously enhance the directional gains for all devices. When $K$ becomes large, the performance advantage of the RA-enabled scheme over the fixed directional and random-orientation schemes gradually diminishes, since the system becomes increasingly dominated by resource sharing and multi-user competition rather than antenna deflection optimization.

\FloatBarrier

\section{Conclusion}
In this letter, we investigated an RA-enabled MEC uplink system for minimizing the maximum computation latency. An efficient AO framework was developed to jointly optimize the MEC computing resource allocation, receive beamforming vectors, and RA deflection angles. Simulation results demonstrated that the proposed RA-enabled MEC design significantly reduces computation latency compared with conventional schemes, showing its potential for low-latency and high-efficiency 6G computing services.

\appendix

\subsection{Derivations of $\nabla u_{k,j}(\vec{\mathbf{f}}_n)$ and $\nabla^2 u_{k,j}(\vec{\mathbf{f}}_n)$}
\label{app:Derivations}
\vspace{-0.1cm}
For the sake of exposition, we reformulated $u_{k,j}(\mathbf{F})$ as
\vspace{-0.15cm}
\small
\begin{equation}
	u_{k,j}(\mathbf F)
	=
	\left|
	\sum_{n=1}^{N} w_{k,n}^*
	\Big(
	\tilde{\beta}_{j,n}(\vec{\mathbf f}_n^T\vec{\mathbf q}_{j,n})^p
	+
	\sqrt{\tfrac{1}{\kappa_j+1}}\tilde g_{j,n}
	\Big)
	\right|^2,
\end{equation}
\normalsize

\vspace{-0.15cm}
To optimize with respect to a specific $\vec{\mathbf{f}}_n$, we fix the other variables $\vec{\mathbf{f}}_m$ for $m \neq n$ and consider $u_{k,j}$ as a function of $\vec{\mathbf{f}}_n$ only. To facilitate the subsequent derivations, we introduce the following auxiliary variables based on $\vec{\mathbf{f}}_{n}$: $b_n \triangleq \vec{\mathbf{f}}_n^T \vec{\mathbf{q}}_{j,n}$, $c_n \triangleq w_{k,n}^* \tilde{\beta}_{j,n}$, $\xi_n \triangleq w_{k,n}^* \sqrt{\frac{1}{\kappa_j+1}} \tilde{g}_{j,n}$, and $d_n \triangleq \sum_{m \neq n} w_{k,m}^* \left( \tilde{\beta}_{j,m} b_m^p + \sqrt{\frac{1}{\kappa_j+1}} \tilde g_{j,m} \right)$.

Thus, the objective function can be written as $u_{k,j}(\vec{\mathbf{f}}_n) = |d_n + \xi_n + c_n b_n^p|^2$. For convenience, we let $\bar{d}_n \triangleq d_n + \xi_n$ and expand this expression, we obtain:
\vspace{-0.1cm}
\small
\begin{equation}
    u_{k,j}(\vec{\mathbf{f}}_n) = |\bar{d}_n|^2 + 2\operatorname{\mathcal{R}}(\bar{d}_n c_n^*) b_n^p + |c_n|^2 b_n^{2p}.
\end{equation}\normalsize

\vspace{-0.1cm}
By defining the real coefficients as $A_n = |\bar{d}_n|^2$, $B_n = 2 \operatorname{\mathcal{R}}(\bar{d}_n c_n^*)$, and $C_n = |c_n|^2$, we arrive at a simplified form:
\vspace{-0.1cm}
\small
\begin{equation}
    \label{eq: Definition of g}
    u_{k,j}(\vec{\mathbf{f}}_n) = A_n + B_n b_n^p + C_n b_n^{2p}.
\end{equation}\normalsize

\vspace{-0.1cm}
\noindent
Taking derivatives of (\ref{eq: Definition of g}) with respect to $b_n$, we obtain:
\vspace{-0.1cm}
\small
\begin{flalign}
&\frac{\partial u_{k,j}(\vec{\mathbf{f}}_n)}{\partial b_n} \!=\! B_n p b_n^{p-1} \!+\! 2C_n p b_n^{2p-1} \!=\! p b_n^{p-1}(B_n \!+\! 2C_n b_n^p),\\
&\frac{\partial^2 u_{k,j}(\vec{\mathbf{f}}_n)}{\partial b_n^2} \!=\! p(p-1) B_n b_n^{p-2} \!+\! 2p(2p-1) C_n b_n^{2p-2}.
\end{flalign}\normalsize

\vspace{-0.1cm}
Applying the chain rule and noting that $\nabla_{\vec{\mathbf{f}}_n} b_n = \vec{\mathbf{q}}_{j,n}$, the gradient vector $\nabla u_{k,j}(\vec{\mathbf{f}}_n)$ and the Hessian matrix $\nabla^2 u_{k,j}(\vec{\mathbf{f}}_n)$ are:
\vspace{-0.1cm}
\small
\begin{flalign}
&\nabla u_{k,j}(\vec{\mathbf{f}}_n) = p b_n^{p-1}(B_n + 2C_n b_n^p)\vec{\mathbf{q}}_{j,n}, &\\
&\nabla^2 u_{k,j}(\vec{\mathbf{f}}_n) = \left[p(p-1) B_n b_n^{p-2} + 2p(2p-1) C_n b_n^{2p-2} \right]\vec{\mathbf{q}}_{j,n}{\vec{\mathbf{q}}_{j,n}}^T. &
\end{flalign}\normalsize

\vspace{-0.3cm}
\subsection{Construction of $\delta_{k,j,n}$}
\label{app:Construction}
\vspace{-0.2cm}
Based on the definition of $\nabla^2 u_{k,j}(\vec{\mathbf{f}}_n)$ in Appendix \ref{app:Derivations}, we have
\vspace{-0.1cm}
\small
\begin{equation}
\Vert \nabla^2 u_{k,j}(\vec{\mathbf{f}}_n) \Vert_2 \le \Vert \nabla^2 u_{k,j}(\vec{\mathbf{f}}_n) \Vert_F. 
\end{equation}\normalsize

\vspace{-0.2cm}
\noindent
Since $\vec{\mathbf{q}}_{j,n}$ is a unit-norm vector, the matrix $\vec{\mathbf{q}}_{j,n}\vec{\mathbf{q}}_{j,n}^T$ is rank-one with a single nonzero eigenvalue equal to 1. Therefore, the spectral norm of $\nabla^2 u_{k,j}(\vec{\mathbf{f}}_n)$ equals the absolute value of its scalar coefficient, which is given by
\vspace{-0.15cm}
\small
\begin{equation}
    \left\Vert \nabla^2 u_{k,j}(\vec{\mathbf{f}}_n) \right\Vert_2 = \left| \frac{\partial^2 u_{k,j}(\vec{\mathbf{f}}_n)}{\partial b_n^2} \right|.
\end{equation}\normalsize

\vspace{-0.1cm}
To ensure a positive semidefinite majorization condition $\delta_{k,j,n} \mathbf{I}_3 \succeq \nabla^2 u_{k,j}(\vec{\mathbf{f}}_n)$, we can directly select $\delta_{k,j,n}$ as
\vspace{-0.05cm}
\small
\begin{equation}
    \delta_{k,j,n} = \left| p(p-1)B_n b_n^{p-2} + 2p(2p-1)C_n b_n^{2p-2} \right|,
\end{equation}\normalsize

\vspace{-0.1cm}
\noindent
which guarantees $\delta_{k,j,n} \mathbf{I}_3 \succeq \nabla^2 u_{k,j}(\vec{\mathbf{f}}_n)$, and hence $\delta_{k,j,n} \ge \Vert \nabla^2 u_{k,j}(\vec{\mathbf{f}}_n) \Vert_2$.
This choice ensures that $\delta_{k,j,n}$ serves as a valid local Lipschitz constant for the gradient mapping, i.e., $\Vert \nabla u_{k,j}(\vec{\mathbf{f}}_n) - \nabla u_{k,j}(\vec{\mathbf{f}}_n^{(i)}) \Vert \le \delta_{k,j,n} \Vert \vec{\mathbf{f}}_n - \vec{\mathbf{f}}_n^{(i)} \Vert$.

\bibliographystyle{IEEEtran}
\bibliography{IEEEabrv,Reference2.bib}

\end{document}